\documentclass[12pt,a4paper]{article}

\usepackage{a4wide}
\usepackage{multirow}
\usepackage{graphicx}
\usepackage{amsmath}
\usepackage{amssymb}
\usepackage{amsfonts}
\usepackage{natbib}
\usepackage{pstricks}
\usepackage{setspace} %%pour gérer efficacement les interlignes
\usepackage{aeguill} %%a virer lors de la soumission

\usepackage[T1]{fontenc}     % pb si pdflatex
\usepackage{psfig}
\usepackage{epsfig}
\usepackage{latexsym}

\renewcommand{\arraystretch}{1.6}

\newenvironment{monitem}[1]
{\begin{list}{#1}
        {
        \setlength{\rightmargin}{\leftmargin}
        \setlength{\itemsep}{-1mm}
        }
}{\end{list}}

\newtheorem{rem}{\bf \it Remark}[section]

\def\ee{I\mskip-7muE} % ESPERANCE
\def\bx{\mbox{\boldmath $x$}}  % a remplacer si on convertit .tex en .rtf

\def\bu{\mbox{\boldmath $u$}}
\def\bv{\mbox{\boldmath $v$}}

\def\bp{\mbox{\boldmath $p$}}
\def\bk{\mbox{\boldmath $k$}}

\def\btheta{\mbox{\boldmath $\theta$}}
\def\bbeta{\mbox{\boldmath $\beta$}}
\def\bSigma{\mbox{\boldmath $\Sigma$}}

\def\bR{\mbox{\boldmath $R$}}
\def\bI{\mbox{\boldmath $I$}}

\begin{document}

\title{An efficient methodology for modeling complex computer codes with Gaussian processes}

\author{Amandine MARREL$^{\ast,1}$, Bertrand IOOSS$^\dag$,\\
Fran\c{c}ois VAN DORPE$^\star$, Elena VOLKOVA$^\ddag$}
\date{}

\maketitle

\begin{center}
%Submitted to: {\it Computational Statistics and Data Analysis}

%\vspace{0.5cm}

$^\ast$ CEA Cadarache, DEN/DTN/SMTM/LMTE

$^\dag$ CEA Cadarache, DEN/DER/SESI/LCFR

$^\star$ CEA Cadarache, DEN/D2S/SPR

$^\ddag$ Kurchatov Institute, Russia

%\vspace{0.5cm}
%\vspace{0.2cm}

\end{center}

\renewcommand{\baselinestretch}{1.4} % double interligne
\renewcommand{\arraystretch}{1.1} % simple interligne pour les tableaux

%\doublespacing
%\twocolumn

\abstract{

Complex computer codes are often too time expensive to be directly used to perform uncertainty propagation studies, global sensitivity analysis or to solve optimization problems. A well known and widely used method to circumvent this inconvenience consists in replacing the complex computer code by a reduced model, called a metamodel, or a response surface that represents the computer code and requires acceptable calculation time. One particular class of metamodels is studied: the Gaussian process model that is characterized by its mean and covariance functions. A specific estimation procedure is developed to adjust a Gaussian process model in complex cases (non linear relations, highly dispersed or discontinuous output, high dimensional input, inadequate sampling designs, etc.). The efficiency of this algorithm is compared to the efficiency of other existing algorithms on an analytical test case. The proposed methodology is also illustrated for the case of a complex hydrogeological computer code, simulating radionuclide transport in groundwater.

}
%\vspace{0.2cm}

\noindent
{\bf Keywords:} Gaussian process, kriging, response surface, uncertainty, covariance, variable selection, computer codes.

\footnotetext[1]{Corresponding author: A. Marrel\\
DEN/DTN/SMTM/LMTE, 13108 Saint Paul lez Durance, Cedex, France\\
Phone: +33 (0)4 42 25 26 52, Fax: +33 (0)4 42 25 62 72\\
Email: amandine.marrel@cea.fr}

\clearpage
%%%%%%%%%%%%%%%%%%%%%%%%%%%%%%%%%%%%%%%%%%%%%%%%%%%%%%%%%%%%%

\section{INTRODUCTION}

With the advent of computing technology and numerical methods, investigation of computer code experiments remains an important challenge.
Complex computer models calculate several output values (scalars or functions) which can depend on a high number of input parameters and physical variables.
These computer models are used to make simulations as well as predictions or sensitivity studies.
Importance measures of each uncertain input variable on the response variability provide guidance to a better understanding of the modeling in order to reduce the response uncertainties most effectively (\cite{salcha00}, \cite{kle97}, \cite{heljoh06}).

However, complex computer codes are often too time expensive to be directly used to conduct uncertainty propagation studies or global sensitivity analysis based on Monte Carlo methods.
To avoid the problem of huge calculation time, it can be useful to replace the complex computer code by a mathematical approximation, called a response surface or a surrogate model or also a metamodel.
The response surface method (\cite{boxdra87}) consists in constructing a function that simulates the behavior of real phenomena in the variation range of the influential parameters, starting from a certain number of experiments.
Similarly to this theory, some methods have been developed to build surrogates for long running computer codes (\cite{sacwel89}, \cite{osiamo96}, \cite{klesar00}, \cite{fanli06}).
Several metamodels are classically used: polynomials, splines, generalized linear models, or learning statistical models such as neural networks, support vector machines, \ldots (\cite{hastib02}, \cite{fanli06}).

For sensitivity analysis and uncertainty propagation, it would be useful to obtain an analytic predictor formula for a metamodel. Indeed, an analytical formula often allows the direct calculation of sensitivity indices or output uncertainties. Moreover, engineers and physicists prefer interpretable models that give some understanding of the simulated physical phenomena and parameter interactions. Some metamodels, such as polynomials (\cite{jouzab04}, \cite{kle05}, \cite{ioovan06}), are easily interpretable but not always very efficient. Others, for instance neural networks (\cite{alamcn04}, \cite{fanli06}), are more efficient but do not provide an analytic predictor formula.

The kriging method (\cite{mat70}, \cite{cre93}) has been developed for spatial interpolation problems; it takes into account spatial statistical structure of the estimated variable.
\cite{sacwel89} have extended the kriging principles to computer experiments by considering the correlation between two responses of a computer code depending on the distance between input variables.
The kriging model (also called Gaussian process model), characterized by its mean and covariance functions, presents several advantages, especially the interpolation and interpretability properties. Moreover, numerous authors (for example, \cite{curmit91}, \cite{sanwil03} and \cite{vazwal05}) show that this model can provide a statistical framework to compute an efficient predictor of code response.

From a practical standpoint, constructing a Gaussian process model implies estimation of several hyperparameters included in the covariance function. This optimization problem is particularly difficult for a model with many inputs and inadequate sampling designs (\cite{fanli06}, \cite{oha06}).
In this paper, a special estimation procedure is developed to fit a Gaussian process model in complex cases (non linear relations, highly dispersed output, high dimensional input, inadequate sampling designs). Our purpose includes developing a procedure for parameter estimation via an essential step of input parameter selection. Note that we do not deal with the design of experiments in computer code simulations (i.e. choosing values of input parameters). Indeed, we work on data obtained in a previous study (the hydrogeological model of \cite{volioo07}) and try to adapt a Gaussian process model as well as possible to a non-optimal sampling design.
In summary, this study presents two main objectives: developing a methodology to implement and adapt a Gaussian process model to complex data while studying its prediction capabilities.

The next section briefly explains the Gaussian process modeling from theoretical expression to predictor formulation and model parameterization. In section 3, a parameter estimation procedure is introduced from the numerical standpoint and a global methodology of Gaussian process modeling implementation is presented. Section 4 is devoted to applications. First, the algorithm efficiency is compared to other algorithms for the example of an analytical test case.
Secondly, the algorithm is applied to the data set ($20$ inputs and $20$ outputs) coming from a hydrogeological transport model based on waterflow and diffusion dispersion equations. The last section provides some possible extensions and concluding remarks.

%***********************************************************

\section{GAUSSIAN PROCESS MODELING}\label{secgpmodel}

\subsection{Theoretical model}

Let us consider $n$ realizations of a computer code. Each realization $y(\bx)$ of the computer code output corresponds to a d-dimensional input vector $\bx = (x_1,...,x_d)$. The $n$ points corresponding to the code runs are called an experimental design and are denoted as $X_s = ( \bx^{(1)},...,\bx^{(n)} )$. The outputs will be denoted as $Y_s = (y^{(1)},...,y^{(n)})$ with $y^{(i)} = y(\bx^{(i)}), i=1,...,n$.
Gaussian process (Gp) modeling treats the deterministic response $y(\bx)$ as a realization of a random function $Y(\bx)$, including a regression part and a centered stochastic process. This model can be written as:
\begin{equation}\label{eqfirst}
 Y ( \bx ) = f ( \bx ) + Z ( \bx) .
\end{equation}

The deterministic function $f(\bx)$ provides the mean approximation of the computer code. Our study is limited to the parametric case where the function $f$ is a linear combination of elementary functions.
%and, more precisely, to the zero, first- and second-degree polynomial regression. This quite simple regression part has been chosen in order to facilitate estimation of the random part contribution.
Under this assumption, $f(\bx)$ can be written as follows:
\[ f ( \bx ) = \sum_{j = 0}^k \beta_j f_j ( \bx ) = F ( \bx ) \bbeta , \]
where $\bbeta = [ \beta_0, \ldots, \beta_k ]\text{}^t $ is the regression parameter vector and
 $F ( \bx ) =  [ f_0 ( \bx ), \ldots, f_k ( \bx ) ]$ is the regression matrix,
with each $f_j$ ($j=0, \ldots, k$) an elementary function. 
In the case of the one-degree polynomial regression, $(d+1)$ elementary functions are used:
 \[
    \left\{ \begin{array}{l}
      f_0 ( \bx ) = 1 ,  \\
      f_i ( \bx ) = x_i \;\text{ for }\; i=1, \ldots,d .
    \end{array}  \right.
  \]

%\begin{rem}
In the following, we use this one-degree polynomial for the regression part, while our methodology can be extended to other bases of regression functions.
The regression part allows the addition of an external drift. Without prior information on the relation between the model output and the input variables, this quite simple choice appears reasonable. Indeed, adding this simple external drift allows for a nonstationary global model even if the stochastic part $Z$ is a stationary process. Moreover, on our tests of section \ref{secappli}, this simple model does not affect our prediction performance. This simplification is also reported by \cite{sacwel89}.
%\end{rem}

The stochastic part $Z(\bx)$ is a Gaussian centered process fully characterized by its covariance function:
$\mbox{Cov} ( Z ( \bx ), Z ( \bu ) ) = \sigma^2 R ( \bx, \bu ),$
where $\sigma^2$ denotes the variance of $Z$ and $R$ is the correlation function that provides interpolation and spatial correlation properties.
To simplify, a stationary process $Z(\bx)$ is considered, which means that correlation between $Z(\bx)$ and $Z(\bu)$ is a function of the difference between $\bx$ and $\bu$. Our study is focused on a particular family of correlation functions that can be written as a product of one-dimensional correlation functions:
\[ \mbox{Cov} ( Z ( \bx ), Z ( \bu ) )  = \sigma^2 R ( \bx - \bu ) = \sigma^2 \prod_{l = 1}^d R_l ( x_l - u_l ) . \]

\begin{figure}[ht]
\begin{center}
\includegraphics[height=8cm,width=12cm]{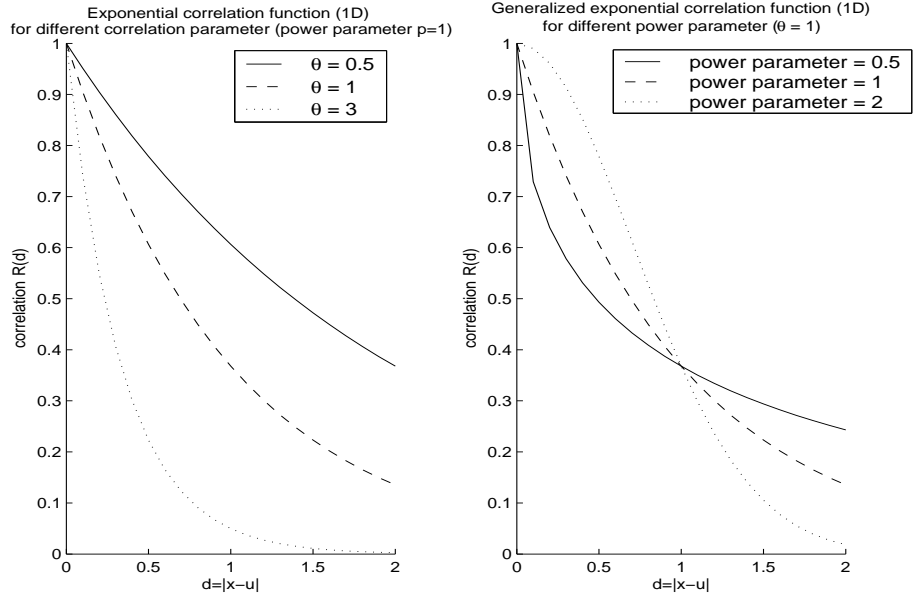}
\caption{Generalized exponential correlation function for different power and correlation parameters.}\label{fig1}
\end{center}
  \end{figure}

\cite{abr94}, \cite{sacwel89}, \cite{chidel99} and \cite{raswil06} give lists of correlation functions with their advantages and drawbacks.
Among all these functions, we choose to use the generalized exponential correlation function:
\[ R_{\btheta, \bp} ( \bx - \bu ) = \prod_{l = 1}^d \exp ( - \theta_l |x_l - u_l |^{p_l} ) \text{ with } \theta_l \geq 0  \text{ and } 0 < p_l \leq 2 ,\]
where  $\btheta = [ \theta_1, \ldots, \theta_d ]\text{}^t $ and $\bp =  [ p_1, \ldots, p_d ]\text{}^t $ are the correlation parameters.
Our motivations stand on the derivation and regularity properties of this function.
Moreover, different choices of covariance parameters allow a wide spectrum of possible shapes (Figure \ref{fig1}); $p = 1$ gives the exponential correlation function and $p=2$ the Gaussian correlation function.

%\begin{rem}\label{rembb}

Even for deterministic computational codes (i.e. outputs corresponding to the same inputs are identical), the outputs may be subject to noise (e.g. numerical noise). In this case, an independent white noise $U(\bx)$ is added in the stochastic part of the model:
\begin{equation}\label{eqfirstnugget}
Y ( \bx ) = f ( \bx ) + Z ( \bx ) + U ( \bx ) ,
\end{equation}
where $U ( \bx ) $ is a centered Gaussian variable with variance $\varepsilon^2 = \sigma^2 \tau$.
In terms of covariance function, this white noise introduces a discontinuity at the origin called the nugget effect (\cite{mat70}):
\[\mbox{Cov} ( Y ( \bx ), Y ( \bu ) ) = \sigma^2 \left( R_{\btheta, \bp} ( \bx - \bu ) + \tau \delta ( \bx-\bu ) \right) ,\]
 where $ \delta ( \bv) = \left\{
    \begin{array}{l}
      1 \mbox{ if } \bv = 0 ,\\
      0 \mbox{ otherwise.}
    \end{array} \right.$
%\end{rem}

  %***********************
\subsection{Joint and conditional distributions} \label{secjoint}

Under the hypothesis of a Gp model, the learning sample $Y_s$ follows the multivariate normal distribution
\[ p( Y_s | X_s ,\bbeta,\sigma,\btheta, \bp, \tau ) = \mathcal{N} \left(F_s \bbeta, \bSigma_s \right) ,\]
where $F_s = [ F ( \bx^{(1)} )\textrm{}^t , \ldots, F ( \bx^{(n)}\textrm{})^t   ]\textrm{}^t $ is the regression matrix and 
\[ \bSigma_s = \sigma^2 \left( R_{\btheta, \bp} \left( \bx^{(i)} - \bx^{(j)} \right)_{i, j = 1 \ldots n} + \tau \bI_n \right) \]
 is the covariance matrix with $\bI_n$ the n-dimensional identity matrix.

If a new point $\bx^{\ast} = (x^{\ast}_1,...,x^{\ast}_d)$ is considered, the joint probability distribution of $(Y_s, Y(\bx^{\ast}))$ is :
\begin{equation}
  p( Y_s, Y(\bx^{*}) | X_s, \bx^{\ast},\bbeta,\sigma,\btheta, \bp, \tau ) = \mathcal{N} \left( \left[ \begin{array}{c}F_s \\ F(\bx^{*}) \end{array}\right] \bbeta,    \left[ \begin{array}{c c} \bSigma_s &  \bk(\bx^{*}) \\ \bk(\bx^{*})\text{}^t & \sigma^2 (1+\tau)  \end{array}\right]  \right) ,
\end{equation}
with
  \begin{equation} 
\begin{array}{lll}
\bk(\bx^{*}) & = &( \: \mbox{Cov}(y^{(1)},Y(\bx^{*})), \ldots, \mbox{Cov}(y^{(n)},Y(\bx^{*})) \: ) \text{}^t  \\
& = & \sigma^2 (  \: R_{\btheta, \bp} (\bx^{(1)},\bx^{*})+ \tau  \delta ( \bx^{(1)}, \bx^{*} ), \ldots, R_{\btheta, \bp} (\bx^{(n)},\bx^{*}) + \tau  \delta ( \bx^{(n)}, \bx^{*} ) \: )  \text{}^t .   
 \end{array} 
\end{equation}  

By conditioning this joint distribution on the learning sample, we can readily obtain the conditional distribution of $Y(\bx^{\ast})$ which is Gaussian (\cite{von64}):
 \begin{equation}
 \begin{array}{l}
  p(Y(\bx^{*}) | Y_s,X_s, \bx^{\ast},\bbeta,\sigma,\btheta, \bp,  \tau ) \\
  = \mathcal{N} \left( \ee [  Y(\bx^{*}) | Y_s,X_s, \bx^{\ast},\bbeta,\sigma,\btheta, \bp,\tau] , \mbox{Var}[  Y(\bx^{*}) | Y_s,X_s, \bx^{\ast} ,\bbeta,\sigma,\btheta, \bp,\tau] \right) ,
\end{array}
\end{equation}
with
 \begin{equation}\label{eq_esperance}  
\left\{ \begin{array}{l}
 \displaystyle \ee [  Y(\bx^{*}) | Y_s,X_s, \bx^{\ast},\bbeta,\sigma,\btheta, \bp,  \tau ] =   F(\bx^{*})\bbeta  +  \bk(\bx^{*})  \text{}^t \bSigma_s^{-1} (Y_s - F_s\bbeta) , \\ 
\displaystyle \mbox{Var}[ Y(\bx^{*}) | Y_s,X_s, \bx^{\ast},\bbeta,\sigma,\btheta, \bp,  \tau ] = \sigma^2 (1+\tau)  -  \bk(\bx^{*}) \text{}^t  \bSigma_s^{-1} \bk(\bx^{*}) .   \end{array}  \right.  
   \end{equation}
   
     The conditional mean (equation (\ref{eq_esperance})) is  used as a predictor. The variance formula corresponds to the mean squared error (MSE) of this predictor and is also known as the kriging variance. This analytical formula for MSE gives a local indicator of the prediction accuracy. More generally, Gp model provides an analytical formula for the distribution of the output variable at an arbitrary new point. This distribution formula can be used for sensitivity and uncertainty analysis, as well as for quantile evaluation (\cite{oha06}). Its use can be completely or partly analytical and avoids costly methods based for example on a Monte Carlo algorithm. The variance expression can also be used in sampling strategies (\cite{schzab04}). All these considerations and possible extensions of Gp modeling represent significant advantages (\cite{curmit91}, \cite{raswil06}).

%***********************
\subsection{Parameter estimation}\label{secestimation}

To compute the mean and variance of a Gp model, estimation of several parameters is needed. Indeed, the Gp model (\ref{eqfirstnugget}) is characterized by the regression parameter vector $\bbeta$, the correlation parameters $(\btheta,\bp)$ and the variance parameters $(\sigma^2,\tau)$.
The maximum likelihood method is commonly used to estimate these parameters. Given a Gp model, the log-likelihood of $Y_s$ can be written as: 
\[ \begin{array}{rl}
l_{Y_s} ( \bbeta, \btheta, \bp, \sigma,\tau ) = & \displaystyle - \frac{n}{2} \ln  ( 2 \pi ) -
  \frac{n}{2} \ln  ( \sigma^2 ) - \frac{1}{2} \ln  (
  \det ( \bR_{\btheta , \bp}+ \tau \bI_n ) ) \\
 & \displaystyle - \frac{1}{2 \sigma^2} ( Y_s - F_s \bbeta ) \text{}^t
  (\bR_{\btheta ,\bp }+ \tau \bI_n )^{- 1} ( Y_s - F_s \bbeta ) . 
  \end{array}\]
Given the correlation parameters $(\btheta,\bp)$ and the variance parameter $\tau$, the maximum likelihood estimator of $\bbeta$ is the generalized least squares estimator:
\begin{equation}\label{eq:betahat}
\hat{\bbeta} = (  F_s \text{}^t  (\bR_{\btheta , \bp}+ \tau \bI_n )^{- 1} F_s )^{- 1}\;\; F_s \text{}^t  (\bR_{\btheta , \bp }+ \tau \bI_n )^{- 1} Y_s ,
\end{equation}
and  the maximum likelihood estimator of $\sigma^2$ is:
\begin{equation}\label{eq:sigma2hat} \widehat{\sigma^2} = \frac{1}{n} ( Y_s - F_s \hat{\bbeta} ) \text{}^t 
( \bR_{\btheta , \bp}+ \tau \bI_n )^{- 1} ( Y_s - F_s \hat{\bbeta} ) .
\end{equation}

\begin{rem}
If we consider the predictor built on the conditional mean (equation (\ref{eq_esperance})), we replace $\bbeta$ by its estimator $\widehat{\bbeta}$. The predictor writes now 
 \[ \widehat{Y(\bx^{*})}_{ | Y_s,X_s, \bx^{\ast},\sigma,\btheta, \bp,  \tau}  =   F(\bx^{*})\widehat{\bbeta}  +  \bk(\bx^{*})  \text{}^t \bSigma_s^{-1} (Y_s - F_s\widehat{\bbeta})
  \]
 and its $MSE$ has consequently an additional component (\cite{sanwil03}): 
 \[
 \mbox{Var}[\widehat{Y(\bx^{*})} | Y_s,X_s, \bx^{\ast},\sigma,\btheta, \bp,  \tau ] =   \sigma^2 (1+\tau)  -  \bk(\bx^{*}) \text{}^t  \bSigma_s^{-1} \bk(\bx^{*}) +  u(\bx^{*}) (F_s \text{}^t \bSigma_s^{-1} F_s)^{-1} u(\bx^{*})\text{}^t 
 \]
with $ u(\bx^{*})  = F(\bx^{*}) -  \bk(\bx^{*})  \text{}^t \bSigma_s^{-1} F_s $.
\end{rem}

Matrix $\bR_{\btheta , \bp}$ depends on $\btheta$ and $\bp$. Consequently, $\hat{\bbeta}$ and $\widehat{\sigma^2}$ depend on $\btheta$, $\bp$ and $\tau$.
Substituting $\hat{\bbeta}$ and $\widehat{\sigma^2}$ into the log-likelihood, we obtain the optimal choice $(\widehat{\btheta},\widehat{\bp},\widehat{\tau})$ which maximizes:
 \[ \phi ( \btheta , \bp, \tau  ) = - \frac{1}{2} \left[ n \ln ( \widehat{\sigma^2} ) + \ln  ( |\bR_{\btheta , \bp}+ \tau \bI_n  | ) \right] \text{
  where    } |\bR_{\btheta , \bp}+ \tau \bI_n  | = \det ( \bR_{\btheta , \bp}+ \tau \bI_n ) .\]
Thus, estimation of $(\btheta,\bp)$ and $\tau$ consists in numerical optimization of the function $\psi$ defined as follows:
\[ \displaystyle (\widehat{\btheta},\widehat{\bp},\widehat{\tau}) = \underset{\btheta , \bp, \tau}{\arg\min} \; \psi ( \btheta , \bp, \tau ) \text{
  with    } \psi ( \btheta , \bp, \tau  ) = |\bR_{\btheta , \bp} + \tau \bI_n |^{\frac{1}{n}}  \: \widehat{\sigma^2} . \]

Our study is focused on complex cases with large dimensions $d$ for the input vector $\bx$ ($d = 20$ in our second example in section \ref{secappli}), where the sampling design has not been chosen as a uniform grid.
In this setting, minimizing function $\psi(\btheta, \bp, \tau )$ is an optimization problem that is numerically costly and hard to solve.
Several difficulties guide the choice of the algorithm.
First, a large number of parameters imposes the use of a sequential algorithm, where different parameters are introduced step by step.
Second, a large parameter domain due to the number of parameters and the lack of prior bounds requires an exploratory algorithm able to explore the domain in an optimal way. Finally, the observed irregularities of $\psi(\btheta, \bp, \tau )$ due, for instance, to a conditioning problem induce local minima, which recommend the use of a stochastic algorithm rather than a descent algorithm.

Several algorithms have been proposed in previous papers.
\cite{welbuc92} use the simplex search method and introduce a kind of forward selection algorithm in which correlation parameters are added step by step to reduce function $\psi(\btheta, \bp, \tau )$.
In Kennedy and O'Hagan's GEM-SA software (\cite{oha06}), which uses the Bayesian formalism, the posterior distribution of hyperparameters is maximized via a conjugate gradient method (the Powel method is used as the numerical recipe).
The DACE Matlab free toolbox (\cite{lopnie02}) introduces a powerful stochastic algorithm  based on the Hooke \& Jeeves method (\cite{bazshe93}), which unfortunately requires a starting point and some bounds to constrain the optimization.
%Moreover, for high dimensional input, this algorithm cannot be applied directly on data including all the input variables.
In complex applications, Welch's algorithm reveals some limitations and for high dimensional input, GEM-SA and DACE software cannot be applied directly on data including all the input variables.
To solve this problem, we propose a sequential version (inspired by Welch's algorithm) of the DACE algorithm.
It is based on the step by step inclusion of input variables (previously sorted).
Our methodology allows progressive parameter estimation by input variables selection both in the regression part and in the covariance function.
The complete description of this methodology is the subject of the next section.

\begin{rem}
 One of the problems we have to acknowledge in the evaluation of $\psi(\btheta, \bp, \tau )$ is the condition number of the prior covariance matrix. This condition number affects the numerical stability of the linear system for the $\hat{\bbeta}$ determination and for the evaluation of the determinant. The degree of ill-conditioning not only depends on sampling design but is also sensitive to the underlying covariance model. \cite{ababag94} showed, for example, that a Gaussian covariance ($p=2$) implies an ill-conditioned covariance matrix (which leads to a numerically unstable system), while an exponential covariance ($p=1$) gives more stability. Moreover, in our case, the experimental design cannot be chosen and numerical parameter estimation is often damaged by the ill-conditioning problem. The nugget effect represented by $\tau$ solves this problem. Although the outputs of the learning sample are no longer interpolated, this nugget effect improves the correlation matrix condition number and increases robustness of our estimation algorithm.
\end{rem}

%***********************************************************

\section{MODELING METHODOLOGY} \label{secmethodo}

Let us first detail the procedure used to validate our model.
Since the Gp predictor is an exact interpolator (except when a nugget effect is included), residuals of the learning data cannot be used directly. So, to estimate the mean squared error in a non-optimistic way, we use either a $K$-fold cross validation procedure (\cite{hastib02}) or a test sample (consisting of new data, unused in the building process of the Gp model).
In both cases, the predictivity coefficient $Q_2$ is computed.
$Q_2$ corresponds to the classical coefficient of determination $R^2$ for a test sample, i.e. for prediction residuals:
 \[ Q_2 ( Y, \hat{Y} ) = 1 - \frac{\sum_{i = 1}^{n_{test}} \left( Y_i - \hat{Y}_i \right)^2}{\sum_{i = 1}^{n_{test}} \left( \bar{Y} - Y_i \right)^2}, \]
  where $Y$ denotes the $n_{test}$ observations of the test set and $\bar{Y}$ is their empirical mean. $\hat{Y}$ represents the Gp model predicted values, i.e. the conditional mean (equation (\ref{eq_esperance})) computed which the estimated values of parameters $(\widehat{\bbeta},\widehat{\sigma},\widehat{\btheta},\widehat{\bp},\widehat{\tau})$.
Other simple validation criteria can be used: the absolute error, the mean and standard deviation of the relative residuals, \ldots (see, for example, \cite{klesar00}), which are all global measures. Some statistical and graphical analyses of residuals can provide more detailed diagnostics.

%\subsection{The general algorithm} \label{secmethodoalgo}

Our methodology consists in seven successive steps.
A formal algorithmic definition is specified for each step.
For $i = 1, \ldots, d$, let $e_i$ denote the $i^{th}$ input variable. $\mathcal{M}_0 = \left\{  e^{(0)}_{1}, \ldots, e^{(0)}_{d} \right\}$ denotes the complete initial model (i.e. all the inputs in their initial ranking). $\mathcal{M}_{1} = \left\{  e^{(1)}_{1}, \ldots, e^{(1)}_{d} \right\}$ and $\mathcal{M}_{2} = \left\{  e^{(2)}_{1}, \ldots, e^{(2)}_{d} \right\}$ refer to the inputs in new rankings after sorting by different criteria (correlation coefficient or variation of $Q_2$). Finally, $\mathcal{M}_{cov}$ and $\mathcal{M}_{reg}$ denote the current covariance model and the current regression model; i.e. the list of selected inputs appearing in the covariance and regression functions.

%\begin{monitem}{$\rhd$}

   \vspace{0.3cm}

{\bf Step 0 - Standardization of input variables}

   \vspace{0.3cm}

\noindent
The appropriate procedure to construct a metamodel requires space filling designs with good optimality and orthogonality properties (\cite{fanli06}).
However, we are not always able to choose the experimental design, especially in industrial studies when the data have been generated a long time ago.
Furthermore, other restrictions can be imposed; for example, a sampling design taking into account the prior distribution of input variables.
This can have prejudicial consequences for hyperparameter estimation and metamodel quality.

So, to increase the robustness of our parameter estimation algorithm and to optimize the metamodel quality, we recommend to transform all the inputs into uniform variables. In order to get each transformed input variable following an uniform distribution $\mathcal{U}[0,1]$, the theoretical distribution (if known) or the empirical ones after a piecewise linear approximation is applied to the original inputs. This approximation is required to avoid transforming a future unsampled $\bx^{*}$ to one of the transformed training sites, even if no element of $\bx^{*}$ is equal to the corresponding element of any of the untransformed training sites.
We empirically observed that this uniform transformation of the inputs seems well adapted to correctly estimate correlation parameters. Choices of bounds and starting points are also simplified by this standardization.

   \vspace{0.3cm}

%  \item
{\bf Step 1 - Initial input variables ranking by decreasing coefficient of correlation between $e_i$ and $Y$}

   \vspace{0.3cm}

\noindent
Sorting input variables is necessary to reduce the number of possible models, especially to dissociate regression and covariance models. Furthermore, direct estimation of all parameters without an efficient starting point gives bad results. So, as a sort criterion, we choose the coefficient of correlation between the input variable and the output variable under consideration.
The correlation coefficients between the input parameters and the output variable are the simplest measures of the influence of inputs on the output (\cite{salcha00}).
They are valid in the linear relation context, while in the nonlinear context, they give a first idea of the hierarchy among input variables, in terms of their influence on the output.
Finally, this simple and intuitive choice does not require any modeling and appears a good initial method to sort the inputs when no other information is available. 

For a strongly nonlinear computer code, it could be interesting to use a qualitative method, independent of the model complexity, in order to sort the inputs by influence order (\cite{heljoh06}). Another possibility would be to fit an initial Gp model with an intercept only regression part and all components of $\bp$ equal to 1 or 2. Only the correlation coefficients vector $\btheta$ has to be estimated.
Then, sensitivity measures such as the Sobol indices (\cite{salcha00}, \cite{volioo07}) are computed and used to sort the inputs by influence order.

\noindent
\underline{Algorithm}

\noindent
  $\mathcal{M}_0= \left\{  e^{(0)}_{1}, \ldots, e^{(0)}_{d} \right\} \Longrightarrow
  \mathcal{M}_{1} = \left\{  e^{(1)}_{1}, \ldots, e^{(1)}_{d} \right\} $\\
  $\left\{ \begin{array}{l}
  \mathcal{M}_{reg} = \mathcal{M}_{1}\\
  \mathcal{M}_{cov} = \mathcal{M}_{1}
   \end{array}  \right.$

 \vspace{0.3cm}

%  \item
\noindent
{\bf Step 2 - Initialization of the correlation parameter bounds and starting points for the estimation procedure}

   \vspace{0.3cm}

To constrain the $\psi$ optimization, the DACE estimation procedure requires three following values for each correlation parameter: a lower bound, an upper bound and a starting point. These values are crucial for the success of the estimation algorithm, when it is used directly for all the input variables. However, using sequential estimation based on progressive introduction of input variables, we limit the problems associated with these three values, especially with the starting point value. Another way to reduce the importance of starting point and bounds is to increase the number of iterations in DACE estimation algorithm. However, in the case of a high number of inputs, increasing the number of iterations in DACE can become extremely time expensive; a compromise has to be found. 
As the input variables have been previously transformed into standardized uniform variables, the initialization and the bounds of the correlation parameters can be the same for all the inputs:
% For the power parameter $p$, $0$, $2$ and $1$ can be respectively taken as lower bound, upper bound and starting value. For the correlation parameter $\btheta$, we finally take $ lob = 10^{-8} $ and  $ upb = 3 $  as bounds and  $\btheta^0  = 0.5 $  as starting value.
 \begin{monitem}{$\Diamond$}
  \item lower bounds for each component of $\btheta$ and $\bp$:
  $lob_{\btheta}   = 10^{-8} $ , $lob_{\bp}   = 0 $,

 \item upper bounds for each component of $\btheta$ and $\bp$:
 $upb_{\btheta}   =  100 $ ,  $upb_{\bp}   =  2 $,

 \item starting points for estimation of each component of $\btheta$ and $\bp$:
 $\theta^0  = 0.5$ , $p^0  = 1$.

 \end{monitem}

  \vspace{0.3cm}

%  \item
\noindent
{\bf Step 3 - Successive inclusion of input variables in the covariance function}

   \vspace{0.3cm}

For each set of inputs included in the covariance function, all the inputs from the ordered set in the regression function are evaluated. Correlation and regression parameters are estimated by the DACE modified algorithm, with the values, estimated at the  ${(i-1)}^{th}$ step for the same regression model, used as a starting point. More precisely, at step $i$, input variables numbered from $1$ to $i$ are included in the covariance function and the algorithm estimates pairs of the correlation parameters $(\theta_l,p_l)$ for $l=1, \ldots, i$. As the starting point, the algorithm uses correlation parameters obtained at the ${(i-1)}^{th}$ step for the starting values of $((\theta_1,p_1),\ldots,(\theta_{i-1},p_{i-1}))$.
First starting value of $(\theta_i,p_i)$ is fixed to an arbitrary reference value. Then, at each step, selection of variables in the regression part is also made.

\cite{hoedav05} recommends the corrected Akaike information criterion (AICC) for input selection in the regression model in order to take spatial correlations into account.
Therefore, after the estimation of correlation and regression parameters, the AICC is computed:
  \[
    \mbox{AICC} = - 2 l_{Y_s} \left( \hat{\bbeta}, \hat{\btheta}, \hat{\sigma}
    \right) + 2 n \frac{m_1 + m_2 + 1}{n - m_1 - m_2 - 2} ,
  \]
  where $m_1$ denotes the number of input variables in the regression function, $m_2$ those in the covariance function and $l_Y$ the log-likelihood of the sample $Y$.
The required model is the one minimizing this criterion.

\noindent
\underline{Algorithm}

\noindent
  For $i = 1 \ldots d$

  \begin{monitem}{$\Diamond$}

  \item Step 3.1: Variables in covariance function

  $\mathcal{M}_{i, cov} =  \mathcal{M}_{cov}( 1, \ldots, i )$

  \item Step 3.2:  Successive inclusion of input variables in regression function

 For $j = 1 \ldots d$
  \begin{itemize}
  \item Regression Model:\\
    $\mathcal{M}_{j, reg}  =  \mathcal{M}_{reg}( 1, \ldots, j )$

  \item Parameter estimation:\\
   $\btheta^{init} = ({\theta_1}^{(i-1),j},\ldots ,{\theta_{i-1}}^{(i-1),j},{\theta}^0) \textrm{}^t $\\
     $\bp^{init} = ({p_1}^{(i-1),j},\ldots ,{p_{i-1}}^{(i-1),j},{p}^0) \textrm{}^t $\\
   $[\theta^{i,j},p^{i,j}] = \textrm{DACE estimation}(\mathcal{M}_{i, cov},\mathcal{M}_{j, reg},[\btheta^{init},\bp^{init}],[lob_{\btheta},lob_{\bp}],[upb_{\btheta},upb_{\bp}]) $

  \item AICC Criterion computation\\
 $\mbox{AICC} (i, j)  =  \mbox{AICC} (\mathcal{M}_{i, cov},\mathcal{M}_{j, reg} ) $

  \end{itemize}
  \noindent
 End
    \item Step 3.3: Optimal regression model selection:\\
  $j^{optim}(i) = \underset{j}{\arg\min \; } ( \mbox{AICC} (i, j))$

  \item Step 3.4: $Q_2$ evaluation by $K$-fold cross validation or on test data (with current correlation model and optimal regression model)\\
  $Q_2(i) = Q_2(\mathcal{M}_{i, cov}, \mathcal{M}_{j^{optim}(i), reg} ) $

  \end{monitem}
  \noindent
  End
  \vspace{0.3cm}

This order (correlation outer, regression inner) can be justified by minimizing the computer time required for optimization. The selection procedure for the regression part is made by the minimization of AICC criterion which requires, at each step, only one parameter estimation. On the other hand, the covariance selection is made by the maximization of $Q_2$ which is often computed by a $K$-fold cross validation. This procedure requires, at each step, $K$ estimation procedures. So, the loop on covariance selection is the more expensive, and consequently has to be outer. The choice of $K$ depends on the number of observations of the data set, on the constraints in term of computer time and on the influence of the learning sample size on prediction quality. If few data are available, a leave-one-out cross-validation could be preferred to a $K$-fold procedure to avoid an undesirably negative effect of small learning sets on prediction quality.

\begin{rem}
To avoid some biases on the choice of the optimal covariance model in the next two steps, the coefficient $Q_2$ has to be computed on a test sample (or by a cross validation procedure), different from the one used for the final validation of the Gp model at step 7.
\end{rem}

Other criteria often used in the optimization of the computer experiment designs (\cite{sacwel89}, \cite{sanwil03}) could be considered to select the optimal regression and covariance model. These criteria are based on the variance of Gp model: they produce a model that minimizes the maximum or the integral of predictive variance over input space. However, in the case of a high number of inputs, the optimization of these criteria can be very computer time expensive. The advantage of the $Q_2$ statistic is its relatively fast evaluation, while producing a final model that optimizes the predictive performance.  

    \vspace{0.3cm}
% \item
\noindent
{\bf Step 4 (optional) - New input variables ranking in the covariance function based on the evolution of $Q_2$ (inputs sorted by decreasing ``jumps'' of $Q_2$)}
% (Sorting new inputs in the covariance function based on the $Q_2$ increase)}

   \vspace{0.3cm}

This optional step improves the selection of inputs, particularly in the covariance function.
For each input $X_i$, the increase of the $Q_2$ coefficient (denoted $\Delta Q_2(i)$) is computed when this  $i^{th}$ variable is added to the covariance function. This value is an indicator of the contribution of the  $i^{th}$ input to the accuracy of the Gp model. For this reason, it can be judicious to use values $\Delta Q_2(1), \ldots,\Delta Q_2(d)$ to sort the inputs included in the correlation function.
The inputs are sorted by decreasing of values $\Delta Q_2(i)$ and the procedure of parameter estimation is repeated with this new ranking of inputs for the covariance function (step 3 is rerun).

\noindent
\underline{Algorithm}

\noindent
  \begin{itemize}
  \item Evaluation of $Q_2$ increase for each input variable included in the covariance function:\\
  $\Delta Q_2(k) = Q_2(1)$\\
  For $k = 2 \ldots d$\\
  \mbox{ } \hspace{0.3cm} $\Delta Q_2(k) = Q_2(k) - Q_2(k-1)$\\
 end

\item
Sorting input variables by decreasing of $\Delta Q_2$ \\
  $ \mathcal{M}_{1} \Longrightarrow \mathcal{M}_{2} $
    \end{itemize}

  \vspace{0.3cm}

%   \item
\noindent
{\bf Step 5 (optional) - Algorithm for parameter estimation with new ranking of input variables in the covariance function}
   \vspace{0.3cm}

This optional step improves the selection of inputs, particularly in the covariance function.
The procedure of parameter estimation (step 3) is repeated with the inputs sorted by decreasing values of $\Delta Q_2(i)$ in the covariance function. Consequently, correlation parameters related to the inputs that are the most influential for the increase of the Gp model accuracy are estimated in the first place. Furthermore, we can also hope that the use of this new ranking allows a decrease in the number of inputs included in the covariance function and an optimal input selection. The use of this new ranking appears more judicious and justifiable for the covariance function than sorting by decreasing correlation coefficient (cf. step 1). However, the ranking of step 1 is kept for the regression function.

\noindent
\underline{Algorithm}

\noindent
  $\left\{ \begin{array}{l}
  \mathcal{M}_{reg} = \mathcal{M}_{1}\\
  \mathcal{M}_{cov} = \mathcal{M}_{2}
   \end{array}  \right.$

    \vspace{0.3cm}

%\item
\noindent
{\bf Step 6 - Optimal covariance model selection}

   \vspace{0.3cm}

For each set of inputs in the covariance function, the optimal regression model is selected based on minimization of the AICC criterion (cf. step 3.3). Then, the predictivity coefficient $Q_2$ is computed either by cross validation or on a test sample (cf. step 3.4).
Finally, the selected covariance model is the one corresponding to the highest $Q_2$ value.

\noindent
\underline{Algorithm}

\noindent
  $i^{optim} = \underset{i}{\arg\max \; }(Q_2(i))$\\
    $\left\{ \begin{array}{l}
  \mathcal{M}_{cov}^{\mbox{\tiny optim}} =  \mathcal{M}_{cov}( 1, \ldots, i^{optim} )  \\
 \mathcal{M}_{reg}^{\mbox{\tiny optim}} = \mathcal{M}_{reg}( 1, \ldots, j^{optim}(i^{optim}) )
   \end{array}  \right.$

   \vspace{0.3cm} 
%\item  
\noindent
{\bf Step 7 - Final validation of the optimal Gp model}
  
   \vspace{0.3cm}
After building and selecting the optimal Gp model, a final validation is necessary to evaluate the predictive performance and to eventually compare it to other metamodels. To do this, coefficient $Q_2$ is evaluated on a new test sample (i.e. data not used in the building procedure). If only few data are available, a cross validation procedure can be considered. So, two cross validation procedures are overlapped; one for building the model and one for its validation. 
%This double loop can be unfortunately time expensive.

\noindent
\underline{Algorithm}

\noindent
  $Q_2^{final} = Q_2(  \mathcal{M}_{cov}^{\mbox{\tiny optim}},\mathcal{M}_{reg}^{\mbox{\tiny optim}} )$

\vspace{0.2cm}
After all the steps of our algorithm (including the step $5$), we can often link the inputs appearing in the covariance and regression functions with the nature of their effects on the output. Indeed, we can generally observed $4$ cases: the inputs with only a linear effect which are supposed to appear only in the regression and excluded from the covariance with the step $5$, the inputs with only a non-linear effect which are excluded from the regression and can then appear in the covariance with the re-ordering of $\mathcal{M}_{cov}$ at step $5$, the inputs with both effects appearing in the regression and covariance functions and, finally, the inactive input variables excluded from both.

%\end{monitem}

%\subsection{Step 0 - Standardization of input variables}

%\subsection{Step 1 - Initial sort of input variables by decreasing correlation coefficient }

%\subsection{Step 2 - Initialization of correlation parameter bounds and starting point in estimation procedure }

%\subsection{Step 3 - Successive inclusion of input variables in covariance function}\label{sec_step3}

%\subsection{Step 4 - Determination of a new ranking for inputs in covariance function, based on the evolution of $Q_2$ }

%\subsection{Step 5 }\label{secoptimselection1}

%\subsection{Step 6 - Optimal model selection}\label{secoptimselection2}

 %***********************************************************
\section{APPLICATIONS} \label{secappli}

\subsection{Analytical test case}

First, an analytical function called the g-function of Sobol is used to illustrate and justify our methodology.
The g-function of Sobol is defined for $d$ inputs uniformly distributed on $\left[ 0,1 \right]^{d}$:
        \[
        g_{\mbox{\tiny Sobol}} (X_1, \ldots, X_d) = \prod _{k=1}^{d} g_k (X_k) \text{ where } g_k (X_k) = \frac{\left| 4X_k - 2 \right| + a_k}{1 + a_k}  \text{ and }
a_k \geq 0.
\]
Because of its complexity (strongly nonlinear and non-monotonic relationship) and the availability of analytical sensitivity indices, the g-function of Sobol is a well known test example in the studies of global sensitivity analysis algorithms (\cite{salcha00}).
The contribution of each input $X_k$ to the variability of the model output is represented by the weighting coefficient $a_k$. The lower this coefficient $a_k$, the more significant the variable $X_k$. For example:
\[ \left\{ \begin{array}{l}
a_k = 0 \rightarrow X_k \text{ is very important,}  \\
a_k = 1 \rightarrow X_k \text{ is relatively important,}\\
a_k = 9 \rightarrow X_k \text{ is non important,}\\
a_k = 99 \rightarrow X_k \text{ is non significant.}
\end{array}  \right. \]
For our analytical test, we choose $a_k = k$.

Applying our methodology to the g-function of Sobol, we illustrate its different steps, especially the importance of rerunning the estimation procedure after sorting the inputs by decreasing $\Delta Q_2$ (cf. steps 4 and 5). At the same time, comparisons are made with other reference software like, for example, the GEM-SA software (\cite{oha06}, freely available at {\it http://www.ctcd.group.shef.ac.uk/gem.html}).%, especially in the case of large dimension $d$ of inputs.

To do this, different dimensions of inputs are considered, from $4$ to $20$: $d = 4, 6, \ldots, 20 $. For each dimension $d$, we generate a learning sample formed by $N_{LS} = d\times10$ simulations of the g-function of Sobol following the Latin Hypercube Sampling (LHS) method (\cite{mckbec79}). Using these learning data, two Gp models are built: one following our methodology and one using the GEM-SA software. For each method, the $Q_2$ coefficient is computed on a test sample of $N_{TS} = 1000$ points. For each dimension $d$, this procedure is repeated $50$ times to obtain an average performance in terms of the prediction capabilities of each method (mean of $Q_2$). The standard deviation of $Q_2$ is also a good indicator of the robustness of each method.
%All these elements so allow to compare the different methods.

For each dimension $d$, the mean and standard deviation of $Q_2$ computed on the test sample using different methods are presented in Table \ref{Sobolperfo}. Three methods are compared: the GEM-SA software, our methodology without steps 4 and 5, and our methodology with steps 4 and 5.

\begin{table}[ht]
\centering
\begin{tabular}{cclcclcclcc}
\hline
\multicolumn{2}{c}{\small g-Sobol} && \multicolumn{2}{c}{GEM-SA software} && \multicolumn{2}{c}{\small Gp methodology} && \multicolumn{2}{c}{\small Gp methodology} \\
\multicolumn{2}{c}{\small simulations} && \multicolumn{2}{c} {} && \multicolumn{2}{c}{ \small without steps 4 and 5} && \multicolumn{2}{c}{\small with steps 4 and 5} \\
\cline{1-2} \cline{4-5} \cline{7-8} \cline{10-11}
&&&&&&&&&& \\
d & $N_{LS}$ && $\overline{Q_2}$ &  $sd$  && $\overline{Q_2}$ &  $sd$  && $\overline{Q_2}$ &  $sd$  \\
\hline
4 & 40 && 0.82  & 0.08 && 0.60 & 0.21 && 0.86 & 0.07 \\
\hline
6 & 60 && 0.67  & 0.24  && 0.59 & 0.16 && 0.85 & 0.05  \\
\hline
8 & 80 && 0.66 & 0.13 &&  0.61 & 0.10 && 0.85 & 0.04  \\
\hline
10 & 100 &&  0.59 & 0.25 &&  0.63 & 0.13 && 0.83 &  0.05 \\
\hline
12 & 120 &&  0.57 & 0.16 && 0.61 & 0.15 && 0.84 & 0.05 \\
\hline
14 & 140 && 0.60  &  0.17 &&  0.61 & 0.14 &&  0.83 & 0.03\\
\hline
16 & 160 && 0.62  & 0.11 &&  0.67 & 0.06 && 0.86 & 0.04 \\
\hline
18 & 180 &&  0.66 & 0.09 && 0.67 & 0.05 && 0.84 & 0.03\\
\hline
20 & 200 &&  0.64 & 0.09 && 0.72 & 0.07 && 0.86 & 0.02\\
\hline
\end{tabular}

\caption{Mean $\overline{Q_2}$ and standard deviation $sd$ of the predictivity coefficient $Q_2$ for several implementations of the g-function of Sobol.}\label{Sobolperfo}

\end{table}

%These results are also taken up in Figure \ref{fig_Sobolperfo}. % Pour moi (Bertrand), cette fuigure n'a pas d'interet
For the values of $d$ higher than 6, our methodology including double selection of inputs (with steps 4 and 5) clearly outperforms the others. More precisely, the pertinence of rerunning the estimation procedure after sorting the inputs by decreasing $\Delta Q_2$ is obvious. The prediction accuracy is much more robust (lower standard deviation of $Q_2$).

The drawback of our methodology lies in the somewhat costly steps 4 and 5.
Indeed, sequential estimation and rerunning of the procedure require many executions of the Hooke \& Jeeves algorithm, particularly in the case of a double cross validation (cf. steps 3.4 and 7 of the algorithm).
Consequently, this approach is much more computer time expensive than the GEM-SA software.
For example, for a simulation with $d = 10$ and $N_{LS} = 100$, the computing time of our approach is on average ten times larger than that of the GEM-SA software.

For a practitioner, a compromise is usually made between the time to obtain the sampling design points and the time to build a metamodel.
As a conclusion of this section, our methodology is interesting for high dimensional input models (more than ten), for inadequate or small sampling designs (a few hundreds) and when simpler methodologies have failed.
%It is worth to use our methodology when the inputs dimension $d$ does not exceed 20 and the size of learning sample some hundreds. 
The  data presented in the next section fall into this scope.%we have to deal with in hydrogeological tranfert code generally respect these conditions as illustrated in the next section.

\begin{rem}
The Gp model used in the GEM-SA software has a gaussian covariance function. 
Our model uses a generalized exponential correlation function even if it requires the estimation of twice as many hyperparameters. Indeed, the sequential approach allows to estimate a large number of hyperparameters.
\end{rem}

\subsection{Application on an hydrogeologic transport code}

Our methodology is now applied to the data obtained from the modeling of strontium 90 (noted $^{90}$Sr) transport in saturated porous media using the MARTHE software (developed by BRGM, the French Geological Survey).
The MARTHE computer code models flow and transport equations in three-dimensional porous formations.
In the context of an environmental impact study, this code is used to model $^{90}$Sr transport in saturated media for a radwaste temporary storage site in Russia (\cite{volioo07}).
One of the final purposes is to determine the short-term evolution of $^{90}$Sr transport in soils in order to help rehabilitation decision making. Only a partial characterization of the site has been made and, consequently, values of the model input parameters are not known precisely.
One of the first goals is to identify the most influential parameters of the computer code in order to improve the characterization of the site in an optimal way.
Because of large computing time of the MARTHE code, \cite{volioo07} propose to construct a metamodel on the basis of the first learning sample.
In the following, our Gp methodology is applied and its results are compared to the previous ones obtained with boosting regression trees and linear regression.

\subsubsection{Data presentation} \label{subsecpres_appli}

Data simulated in this study are composed of $300$ observations. Each simulation consists of $20$ inputs and $20$ outputs.
The $20$ uncertain model parameters are permeability of different geological layers composing the simulated field (parameters 1 to 7), longitudinal dispersivity coefficients (parameters 8 to 10), transverse dispersivity coefficients (parameters 11 to 13), sorption coefficients (parameters 14 to 16), porosity (parameter 17) and meteoric water infiltration intensities (parameters 18 to 20). To study sensitivity of the MARTHE code to these parameters, simulations of these $20$ parameters have been made by the LHS method. 

For each simulated set of parameters, MARTHE code computes transport equations of $^{90}$Sr and predicts the evolution of $^{90}$Sr concentration. 
Initial and boundary conditions for the flow and transport models are fixed at the same values for all simulations.
% (initial distribution of $^{90}$Sr concentration is shown on Figure \ref{figconcentrations} (a)).
So, for an initial map of $^{90}$Sr concentration in 2002 and a set of 20 input parameter values, MARTHE code computes a map of predicted concentrations in 2010. For each simulation, the 20 outputs considered are values of $^{90}$Sr concentration, predicted for year 2010, in $20$ piezometers located on the waste repository site.
%(an example of resulting $^{90}$Sr concentration map in 2010 is shown on Figure \ref{figconcentrations} (b)).

\subsubsection{Comparison of three different models}

For each output, we choose to compare and analyze the results of three models:
\begin{monitem}{$\triangleright$}
\item Linear regression: it represents a model that provides a reference for the contribution of the Gp model stochastic component to modeling quality. Indeed, comparison between simple linear regression and Gp model will show if considering spatial correlations has significant impact on the modeling results.
Moreover, a selection based on the AICC criterion is implemented to optimize the results of the linear regression.

\item Boosting of regression trees: this model was used in the previous study of the data (\cite{volioo07}).
The boosting trees method is based on sequential construction of weak models (here regression trees with low interaction depth), that are then aggregated.
The MART algorithm (Multiple Additive Regression Trees), described in \cite{hastib02}, is used here.
The boosting trees method is relatively complex, in the sense that, as with neural networks, it is a black box model, efficient but quite difficult to interpret.
It is interesting to see if a Gp model, that is easier to interpret and offers a quickly computable predictor, can compete with a more complex method in terms of modeling and prediction quality.
Note that the boosting trees algorithm also makes its proper input selection.

\item Gaussian process: to implement this model, the methodology previously described in this paper is applied, with the input selection procedure.
\end{monitem}

\subsubsection{Results}

To compare prediction quality of the three different models presented above, the coefficient of predictivity $Q_2$ is estimated by a $6$-fold cross validation. Note that for each model the results correspond to the optimal set of inputs included in the model.
To avoid some bias in the results, the cross validation used to select variables in the Gp model (see step 6) differs from the cross validation used to validate and compare prediction capabilities of the three models. Indeed, at each cross validation step (used to validate), data are divided into a learning sample (denoted $LS1$) of $250$ observations and a test sample ($TS1$) of the $50$ remaining observations.
For the Gp model, the procedure of variable selection is then performed by a second cross validation on $LS1$ (for example: a $4$-fold cross validation, dividing $LS1$ into a learning set $LS2$ of $210$ data and a test set $TS2$ of the 40 others).
Then, an optimal set of variables is determined and a Gp model is built based on the $250$ data of $LS1$ (with this optimal set of inputs previously selected). Finally, the model is validated on the test set $TS1$ that has never been used for the Gp model construction.

\begin{figure}[!ht]
\begin{center}
\includegraphics[height=16.cm,width=9.cm,angle=90]{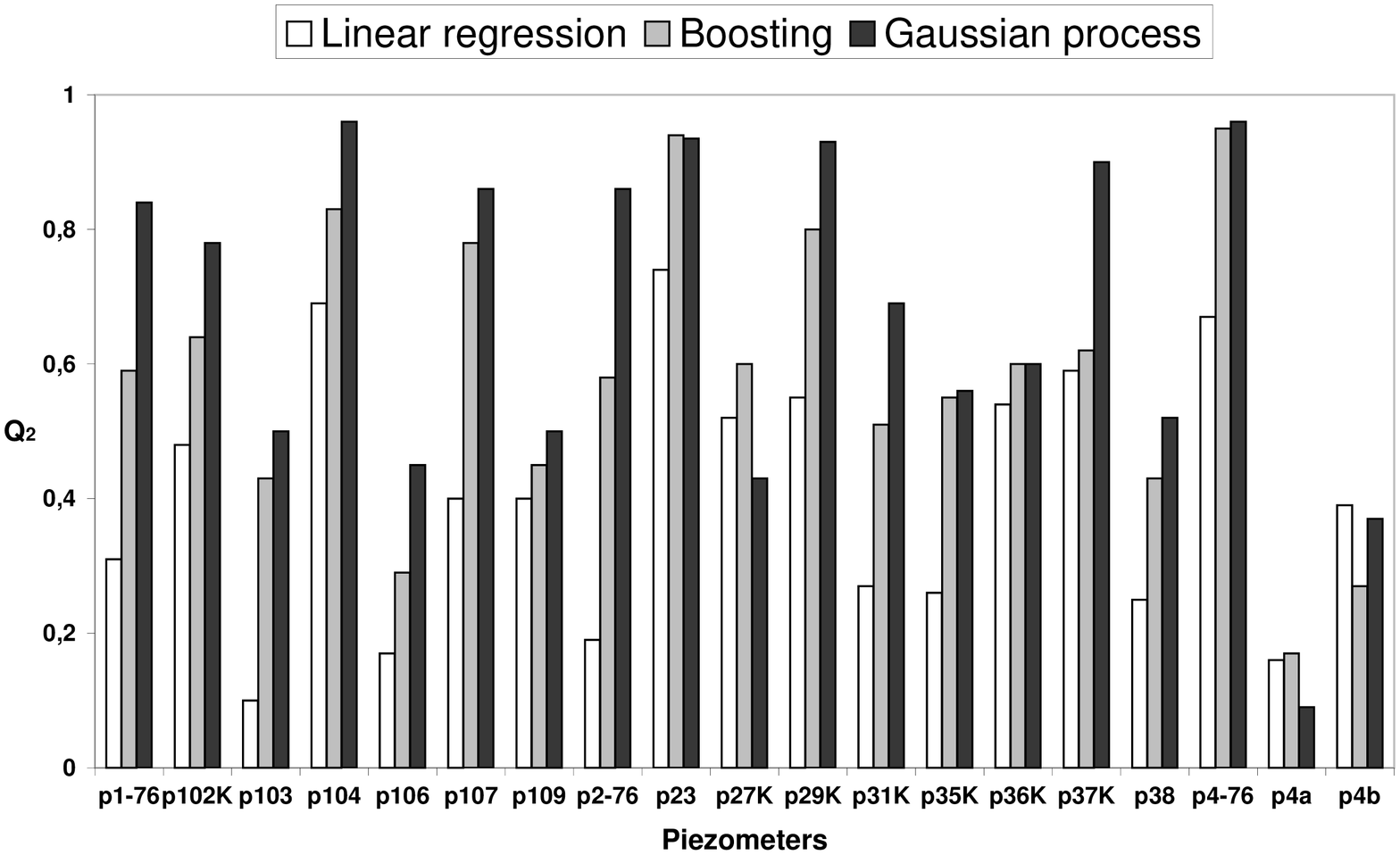}

\vspace{-1cm}
\caption{Barplot of the predictivity coefficient $Q_2$ for the three different models.}\label{elenabarplot}
\end{center}
  \end{figure}

The results are presented in Table \ref{elenaperfo} and are taken up in a barplot (see Figure \ref{elenabarplot}).
Results obtained for the output 8 (piezometer p110) are not considered because of physically insignificant concentration values.
For most outputs, the Gp performance is superior to linear regression and boosting, in many cases substantially so.
Concerning the outputs 11 ($p27k$) and 19 ($p4a$), the performances of the Gp model are worse than the linear regression ones.
However, for these two outputs, the prediction errors are very high and consequently  the difference of performance between the two models can be considered as non-significant.
%do not appear in the table because the concentration values computed by MARTHE are close to zero and thus considered as not physically significant.

\begin{table}[ht]
\begin{center}
\begin{tabular}{cclccc}

\hline
\multicolumn{2}{c}{Output} && Linear regression & boosting trees &  Gaussian process \\
\cline{1-2}
Denomination & Number &&  $Q_2$ &  $Q_2$  &  $Q_2$ \\
\hline
p1-76 & 1 &&  0.31 & 0.59  & 0.84 \\
\hline
p102K & 2 && 0.48 & 0.64  & 0.78\\
\hline
p103 & 3 && 0.10 & 0.43  & 0.5 \\
\hline
p104 & 4  && 0.69 & 0.83 & 0.96 \\
\hline
p106 & 5 && 0.17  & 0.29 &  0.45 \\
\hline
p107 & 6 && 0.40 & 0.78 & 0.86 \\
\hline
p109 & 7 && 0.40 & 0.45 & 0.5 \\
\hline
p2-76 & 9 && 0.19 & 0.58  &  0.86\\
\hline
p23 & 10 && 0.74  & 0.94 & 0.935 \\
\hline
p27K & 11  && 0.52 & 0.60 &  0.43 \\
\hline
p29K & 12  && 0.55 & 0.80 & 0.93  \\
\hline
p31K & 13  && 0.27 &  0.51 & 0.69  \\
\hline
p35K & 14  && 0.26  &  0.55 & 0.56  \\
\hline
p36K & 15 && 0.54  & 0.60  & 0.60 \\
\hline
p37K & 16  && 0.59 & 0.62   & 0.90  \\
\hline
p38 & 17 && 0.25  & 0.43 & 0.52  \\
\hline
p4-76 & 18 && 0.67   & 0.95 & 0.96 \\
\hline
p4a & 19 && 0.16  & 0.17 & 0.09 \\
\hline
p4b & 20 && 0.39 & 0.27  & 0.37  \\
\hline
\end{tabular}
\caption{Predictivity coefficients $Q_2$ for the three different models of the MARTHE data.}\label{elenaperfo}
\end{center}
\end{table}

\begin{figure}[!ht]
     \begin{center}
\psfig{figure=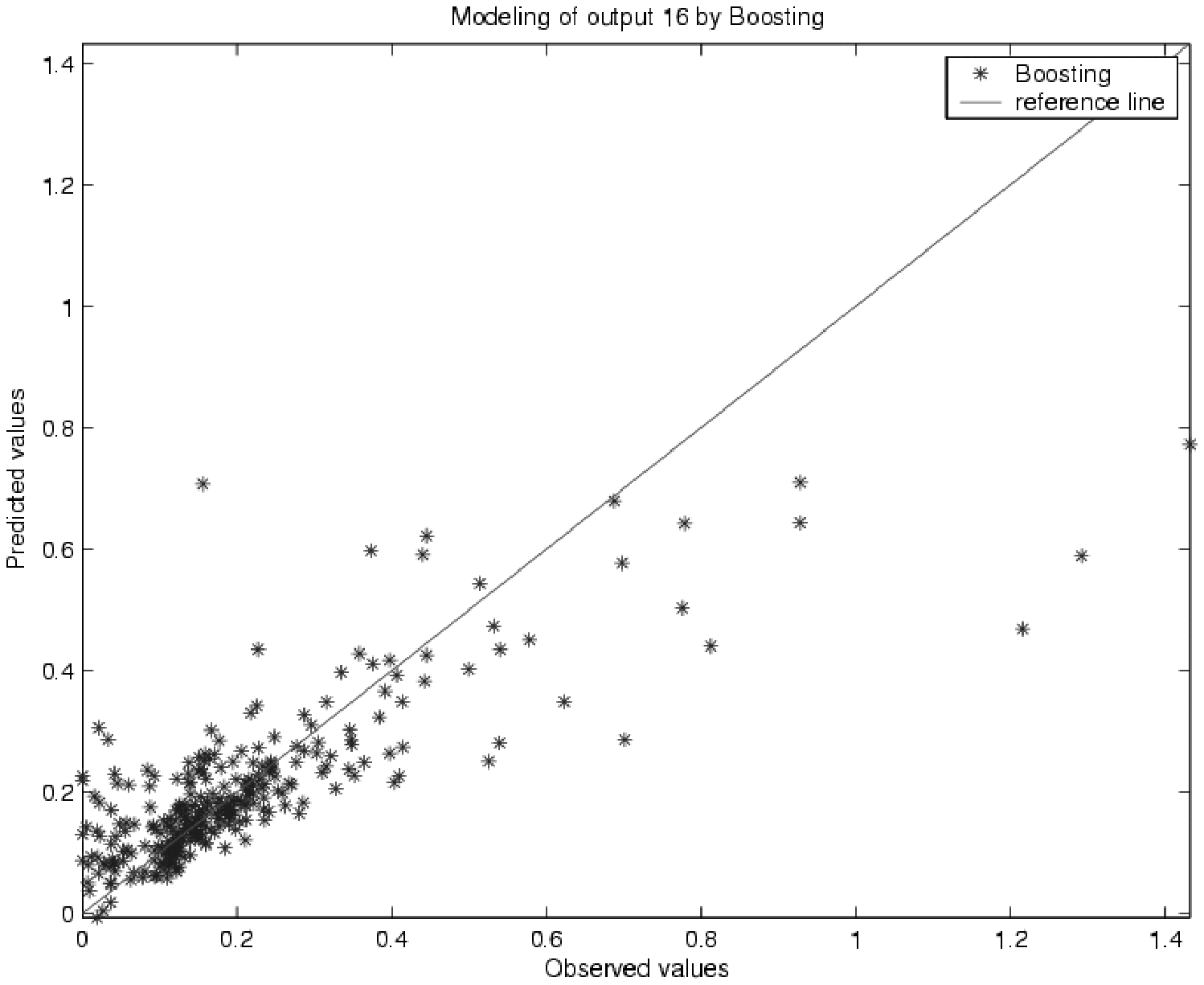,height=7.cm,width=7cm}
\hspace{1cm} \psfig{figure=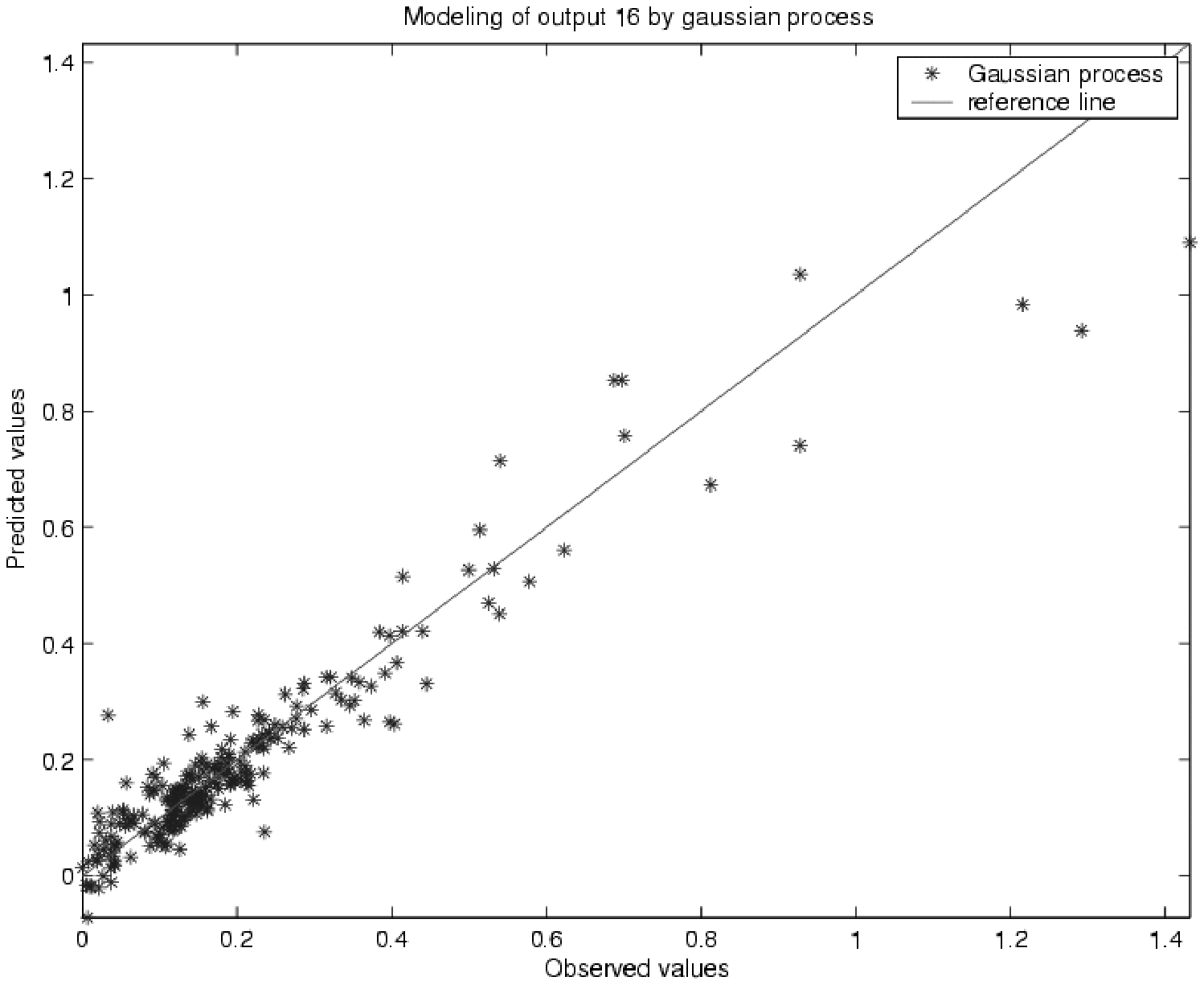,height=7.cm,width=7cm}
\caption{Plot of predicted values vs real values for boosting trees (left) and Gaussian process (right).}\label{figresidu16}
\end{center}
  \end{figure}

As expected, for most of the outputs, the linear regression presents the worst results. When this model is successfully adapted, the two others are also efficient.
When linear regression fails (for example, for output number 12), Gp model presents a real interest, since it gives results as good as those of the boosting trees method. In fact, this is verified for all the ouputs and results are significantly better for several outputs (outputs 1, 2, 4, 9, 12, 13 and 16). To illustrate this, the Figure \ref{figresidu16} shows the predicted values vs real values for the output 16, for the Gaussian process and boosting trees models.
It clearly shows a better repartition of the Gp model residuals than the boosting trees model ones.

\begin{figure}[!ht]
\begin{center}
\includegraphics[height=8.cm,width=12.cm]{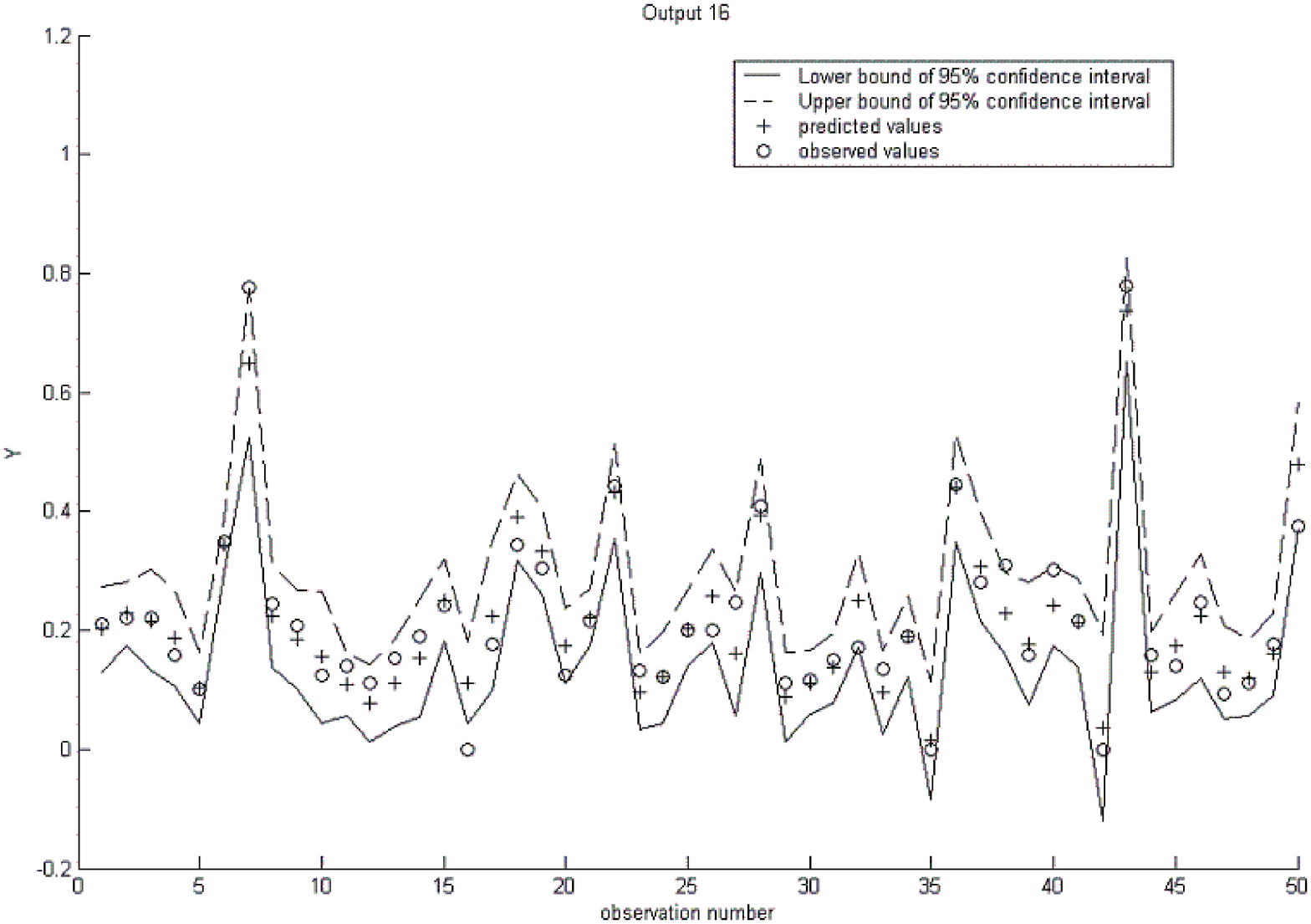}
\caption{Plot of observed and Gaussian process predicted values for the output 16 with the $95\%$ prediction interval based on $\widehat{MSE}$ formula.}\label{figMSE16}
\end{center}
  \end{figure}

Furthermore, the estimator of MSE, that is expressed analytically (see Equation (\ref{eq_esperance})), can be used as a local prediction interval. To illustrate this, we consider 50 observations of the output 16.
Figure \ref{figMSE16} shows the observed values, the predicted values and the upper and lower bounds of the 95\% prediction interval based on the MSE local estimator.
It confirms the good adequacy of the Gp model for this output because all the observed values (except one point) are inside the prediction interval curves.

\subsubsection{Analysis}

These results confirm the potential of the Gp model and justify its application for computer codes.
Application of our methodology to complex data also confirms the efficiency of our input selection procedure. For a fixed set of inputs in the covariance function, we can verify that this procedure selects the best set of inputs in regression part.
Furthermore, the necessity of conducting sequential and ordered procedure estimation has been demonstrated. Indeed, if all the Gp parameters (i.e. considering the 20 inputs) are directly and simultaneously estimated with the DACE algorithm, they are not correctly determined and poor results in terms of $Q_2$ are obtained.
So, in case of a complex model with a large number of inputs, we recommend using a selection procedure such as the algorithm of section \ref{secmethodo}.

The study of these data have motivated the choice of this methodology.
At first, Welch's algorithm (see section \ref{secestimation}) has been tried.
Considering the poor results obtained, our methodology based on the DACE estimation algorithm has been developed. 
To illustrate this, let us detail the different results obtained on the output number 9. With our methodology based on the DACE estimation, the $Q_2$ coefficient (always computed by a $6$-fold cross validation) is $0.86$, while with Welch's algorithm (used in its basic version), $Q_2$ is close to zero.
The difference in the results between the two methods can be explained by the value of estimated correlation parameters which are significantly different.
%Even if the Welch algorithm is very fast, it seems to be not efficient for such kind of real and complex data.

%\begin{rem}
To minimize the number of correlation parameters and consequently reduce computer time required for estimation, the possible values of power parameters $p_i$ ($i=1,\ldots,d$) can be limited to $0.5$, $1$ and $2$. It can be a solution to optimize computer time. It allows an exhaustive, quick and optimal representation of different kinds of correlation functions (two kinds of inflexion are represented). Furthermore, in many cases, estimation of power parameter with generalized exponential correlation converges to exponential ($p_i = 1$) or Gaussian ($p_i = 2$) correlation.
%\end{rem}

%***********************************************************

\section{CONCLUSION} \label{secdisc}

The Gaussian process model presents some real advantages compared to other metamodels: exact interpolation property, simple analytical formulations of the predictor, availability of the mean squared error of the predictions and the proved efficiency of the model.
The keen interest in this method is testified by the publication of the recent monographs of \cite{sanwil03}, \cite{fanli06} and \cite{raswil06}.

However, for its application to complex industrial problems, developing a robust implementation methodology is required.
In this paper, we have outlined some difficulties arising from the parameter estimation procedure (instability, high number of parameters) and the necessity of a progressive model construction.
Moreover, an a priori choice of regression function and, more important, of covariance function is essential to parameterize the Gaussian process model.
The generalized exponential covariance function appears in our experience as a judicious and recommended choice.
However, this covariance function requires the estimation of $2d$ correlation parameters, where $d$ is the input space dimension. In this case, the sequential estimation and selection procedures of our methodology are more appropriate. 
This methodology is interesting when the computer model is rather complex (non linearities, threshold effects, etc.), with high dimensional inputs ($d > 10$) and for small size samples (a few hundreds).

Results obtained on the MARTHE computer code are very encouraging and place the Gaussian process as a good and judicious alternative to efficient but non-explicit and complex methods such as boosting trees or neural networks.
It has the advantage  of being easily evaluated on a new parameter set, independently of the metamodel complexity.
Moreover, several statistical tools are available because of the analytical formulation of the Gaussian model. For example, the MSE estimator offers a good indicator of the model's local accuracy. In the same way, inference studies can be developed on parameter estimators and on the choice of the experimental input design.
Finally, one possible improvement in our construction algorithm is based on the sequential approach of the choice of input design, which remains an active research domain (\cite{schzab04} for example).

%Besides continuing to improve parameter estimation procedure and develop input selection, other extensions of the Gaussian process model can be envisaged. One of these would be cokriging, i.e. simultaneous Gaussian process model of several outputs. It considers not only covariance function of each output, but also cross covariance function between outputs. For example, in the MARTHE data the $20$ outputs correspond to $^{90}$Sr concentration at the $20$ piezometers distributed on the site. Consequently, these outputs can be logically considered as spatially correlated (we refer here to the 2D physical space). So, it appears interesting to model simultaneously several outputs considering a cross covariance function.
%Another possible extension is modeling the concentration over time. Indeed, we only considered the concentration at a fixed moment of time (December of 2010), while the values computed from 2002 to 2010 are also available. Instead of modeling only a scalar function, it would be interesting to use a Gaussian process to represent a transient function. So, extending Gaussian process model to a spatiotemporal function represents the next step of this work.

%%%%%%%%%%%%%%%%%%%%%%%%%%%%%%%%%%%%%%%%%%%%%%%%%%%%%%%%%%%%%

\section*{ACKNOWLEGMENTS}     %>>>> equivalent to \section*{ACKNOWLEDGMENTS}

This work was supported by the MRIMP project of the ``Risk Control Domain'' that is managed by CEA/Nuclear Energy Division/Nuclear Development and Innovation Division.
We are grateful to the two referees for their comments which significantly improved the paper.

%***********************************************************
%\section*{REFERENCES}

% Bibliographie
\singlespacing
\bibliographystyle{elsart-harv}
\bibliography{bibl_hal}

\clearpage
\pagestyle{empty}

\end{document}